\documentstyle[12pt,fullpage,subeqn]{article}
\def\be{\begin{equation}}
\def\ee{\end{equation}}
\def\beq{\begin{eqnarray}}
\def\eeq{\end{eqnarray}}
\def\bea{\begin{array}}
\def\eea{\end{array}}
\def\eq#1{(\ref{#1})}
\def\Large{\large}

\renewcommand{\Large}{\large}
\catcode `@=11 \@addtoreset{equation}{section} \catcode `@=12
\begin{document}
\begin{flushright}
CERN-TH/95-69\\
TIFR/TH/95-11\\
hep-th/9503172
\end{flushright}
\bigskip
\begin{center}
{\Large{\bf STRING BETA FUNCTION EQUATIONS \\[2mm]
FROM THE $c=1$  MATRIX MODEL
}} \\
\bigskip
{\large Avinash Dhar}\footnote{On leave from Theoretical Physics
Group, Tata Institute of Fundamental Research, Homi Bhabha Road,
Bombay 400 005, India.},\footnote{e-mail: adhar@surya11.cern.ch} \\
Theory Division, CERN, CH-1211, Geneva 23, SWITZERLAND \\
\bigskip
{\large Gautam Mandal}\footnote{e-mail: mandal@theory.tifr.res.in}
$\;$ {\large and} $\;$
{\large Spenta R. Wadia}\footnote{e-mail: wadia@theory.tifr.res.in} \\
Tata Institute of Fundamental Research, \\ Homi Bhabha Road,
Bombay 400 005, INDIA
\end{center}
\abstract{We derive the $\sigma$-model tachyon $\beta$-function
equation of 2-dimensional string theory, in the background of flat
space and linear dilaton, working entirely within the $c=1$ matrix
model.  The tachyon $\beta$-function equation is satisfied by a
\underbar{nonlocal} and \underbar{nonlinear} combination of the
(massless) scalar field of the matrix model.  We discuss the
possibility of describing the `discrete states' as well as other
possible gravitational and higher tensor backgrounds of 2-dimensional
string theory within the $c=1$ matrix model.  We also comment on the
realization of the $W$-infinity symmetry of the matrix model in the
string theory.  The present work reinforces the viewpoint that a
nonlocal (and nonlinear) transform is required to extract the
space-time physics of 2-dimensional string theory from the $c=1$
matrix model.
}

\newpage

\setcounter{section}{-1}
\section{Introduction}

Two-dimensional string theory bears, in many ways, the same
relationship to its higher dimensional counterparts as do
low-dimensional, exactly solvable field theory models to their less
tractable counterparts in 4 dimensions. It is the `simplest' string
theory one can imagine; it has a massless propagating mode and
vestiges of the massive string modes in the form of `discrete states'
\cite{one,two}; it possesses a very large symmetry group $W_\infty$
[3-7]; it has a black hole solution to the classical beta-function
equations \cite{nine}; and most remarkably it also has a
nonperturbative formulation in terms of an integrable theory of
noninteracting nonrelativistic fermions of the $c = 1$ matrix model
[9-11,24]. The last feature makes it an
ideal testing ground for discussing issues of strong coupling string
theory and issues related to black hole evaporation and gravitational
collapse.

To address these questions of string theory in the matrix model,
however, we first need a detailed mapping between the two.  This is a
nontrivial problem since the natural `space-time' parameters of the
matrix model \underbar{do not} have the interpretation of the
space-time variables of the string theory. This arises as a
consequence of the fact \cite{two,fourteen}\ that one has to use the
so-called `leg-pole' transformation to the asymptotic wave-functions
to arrive at the S-matrix of the string tachyon from that of the
matrix model scalar.  The `leg-pole' prescription in position space
corresponds to a nonlocal transform of the asymptotic
wave-functions. Using this nonlocal transform it has been shown in
\cite{thirteen}\ that the S-matrix of the matrix model actually
reproduces that of the string theory, which comes from a target space
action in the background of flat space and linear dilaton
\footnote{The general idea of non-local transforms has previously
appeared in the context of mapping of matrix model to string theory in
other backgrounds, {\it e.g.} in [12-14].}.

In this work we will provide further evidence to reinforce the
viewpoint that a nonlocal and, as we shall see, nonlinear mapping is
required to extract the space-time physics of the string theory from
the matrix model. More specifically, we will show that a specific
nonlocal and nonlinear combination of the scalar field of the matrix
model satisfies the $\sigma$-model tachyon $\beta$-function equation
of 2-dimensional string theory \cite{fifteen, sixteen} in the
background of flat space and linear dilaton. Moreover, the
quantization inherited from the matrix model implies a canonical
quantization for this particular combination, thus completing its
identification with the tachyon of string theory (in flat
background). We will discuss the possibility of describing the
discrete states as well as other backgrounds, including those of
higher tensor modes of the string, within the existing formulation of
the matrix model, by employing similar transforms.  Finally, we will
show that the $W$-infinity symmetry of the matrix model has a
\underbar{nonlocal} realization in the string theory, a result that
was earlier derived using the methods of BRST cohomology in liouville
string theory \cite{seven}. Throughout this paper we will be working
in the framework of perturbation theory. For instance we will not
attempt to generalize our mapping to situations in which the fermi
fluid goes over to the other side of the potential barrier.

\section{Review of Some Aspects of the Matrix Model}

Here we will briefly summarize some aspects of the $c=1$ matrix model
that will be relevant to the discussion in the following sections.
For a more extended review which covers this, see \cite{view}.

In the double-scaling limit the $c=1$ matrix model is mapped to a model
of noninteracting nonrelativistic fermions in an inverted harmonic
oscillator potential \cite{eight}\ in one space dimension. The
single-particle hamiltonian for this model is
\be
\label{1}
h(p,q) = {1 \over 2} \left(p^2 - q^2\right)
\ee
where $(p, q)$ label the single-particle phase space of the fermions.
There is a convenient (for many calculations in the matrix model)
field theoretic description for the double-scaled model in terms of
free nonrelativistic fermions \cite{nineteen}. The fermion field,
which we denote by $\psi (q, t)$, satisfies the equation of motion
\be
\label{2}
i\partial_t \psi (q, t) = -~ {1 \over 2} \left(\partial^2_q + q^2
\right) \psi (q, t)
\ee
and its conjugate $\psi^\dagger (q, t)$ satisfies the complex
conjugate of eqn. \eq{2}. The ground state of this model is the fermi
vacuum obtained by filling up to the energy level $\mu$ $(< 0)$. The
semiclassical limit is obtained as $|\mu| \rightarrow \infty$ and in
this limit the fermi surface is described by the hyperbola
\be
\label{3}
{1 \over 2} \left(p^2 - q^2 \right) = \mu = - |\mu| \;.
\ee

The basic building block for the present work will be the phase space
density of fermions, which we denote by $u (p, q, t)$. In terms of the
fermi field $\psi (q, t)$ it is defined as
\be
\label{4}
u(p, q, t) \equiv \int d \lambda e^{-ip\lambda} \psi^\dagger
\left(q - {\lambda
\over 2}, t\right) \psi \left(q + {\lambda \over 2}, t \right)
\ee
and it satisfies the equation of motion
\be
\label{5}
\left(\partial_t + p\partial_q + q\partial_p \right) u (p, q, t) = 0
\; ,
\ee
which follows from eqn. \eq{2}, or by directly using the hamiltonian
\be
\label{6}
H = \int {dpdq \over 2\pi} h (p, q) u (p, q, t)
\ee
and the equal-time commutation relation for the phase space density
$u (p, q, t)$, which follows from its definition, eqn. \eq{4}, in
terms of underlying fermions:
\beq
\label{7}
\left[u(p,q,t), u(p',q',t)\right] &=& -4 \int {dp''dq''
\over 2\pi} u(p'',q'',t) \nonumber \\[2mm]
& & \left[\exp 2i \{p(q'-q'') + p'(q''-q)
+ p''(q-q')\} - c.c. \right]
\eeq

Equation \eq{7} is also a version of the large symmetry algebra, the
$W$-infinity algebra \cite{twenty}, which is a symmetry of the matrix
model [3-7].
The more standard version of the generators of this symmetry algebra
is the following:
\be
\label{8}
W_{mn} = e^{-(m-n)t} \int {dpdq \over 2\pi} (-p-q)^m (p-q)^n u (p,q,t),
\ee
where $m, n \geq 0$. One can easily check, using eqn. \eq{5}, that
$W_{mn}$ are conserved. They satisfy the classical algebra
\be
\label{9}
\{W_{mn}, W_{m'n'}\} = 2 (m'n - mn') W_{m+m'-1, n+n'-1}.
\ee
The quantum version of this is more complicated, but can be computed
using eqn. \eq{7}.

The above phase space density formalism was first introduced in the
present context in \cite{twentyone} and using this variable a
bosonization of the model was carried out \cite{twentyone, twentytwo}.
A crucial ingredient in that bosonization is a quadratic constraint
satisfied by $u(p,q,t)$ \cite{twentyone}. In the semi-classical
limit this quantum constraint reduces to the simpler equation
\be
\label{10}
u^2 (p,q,t) = u(p,q,t) \;.
\ee
Moreover, one also has the constraint of fixed fermion number, which
implies that fluctuations above the fermi surface, eqn. \eq{3},
satisfy
\be
\label{11}
\int {dpdq \over 2\pi} \delta u(p,q,t) = 0, \; \quad \delta
u(p,q,t) \equiv u(p,q,t) - u_0(p,q)
\ee
where $u_0(p,q)$ describes the fermi vacuum. In this way we recover
the Thomas-Fermi limit of an incompressible fermi fluid. The dynamics
of the fluctuations $\delta u(p,q,t)$, which satisfies eqn. \eq{11}
and another constraint because of eqn. \eq{10}, resides only in the
boundary of the fermi fluid (in the semi-classical limit that we are
considering here) \cite{twentythree,twentythreea}.

In the following sections we will use the general framework developed
above.  It turns out that we never need to solve the constraints,
eqns. \eq{10} and \eq{11}, and introduce an explicit parametrization
of the fluid boundary fluctuations $\delta u(p,q,t)$.  In fact, we
will be able to develop the entire formalism treating `$p$' and `$q$'
more or less symmetrically.  This is important since, as we shall see,
in this way we are able to avoid spurious singularities, such as the fold
singularity \cite{polreview}, which may be dynamically generated in an
otherwise perfectly nonsingular initial parametrization of the fluid
boundary fluctuation.  We are able to avoid these singularities
because extracting space-time physics from the matrix model requires a
nonlocal transform.  In a sense, therefore, this is a bonus of the
necessity of a mapping from the matrix model to string theory.

Although our general discussion will never need an explicit
parametrization of $\delta u(p,q,t)$, it will, nevertheless, be useful
at times to express things in a familiar parametrization of the
fluctuations.  For this reason we now summarize, in the rest of this
section, some relevant aspects of the `quadratic profile'
\cite{twentythree,twentythreea} or `collective field' \cite{twentyfour}
parametrization of the fluctuations $\delta u(p,q,t)$.

In the semiclassical limit the fermi vacuum is described by the
density
\be
\label{12}
u_0 (p,q) = \theta (P^0_+(q) - p) \theta(p - P^0_-(q)),
\ee
where
\be
\label{13}
P^0_\pm (q) \equiv \pm P_0 (q) = \pm \sqrt{q^2 + 2\mu}
\ee
satisfy the equation that describes the fermi surface hyperbola, eqn.
\eq{3}.  The quadratic profile or collective field approximation
corresponds to a description of small ripples on the fermi surface by
a density of the form
\be
\label{14}
u(p,q,t) = \theta(P_+(q,t) - p) \theta(p - P_-(q,t)).
\ee
Substituting this in eqn. \eq{5}, we get the equations of
motion of $P_\pm$:
\be
\label{15}
\partial_t P_\pm (q,t) = \frac{1}{2} \partial_q (q^2 - P^2_\pm (q,t)).
\ee
This equation is clearly solved by the fermi vacuum, eqns. \eq{12} and
\eq{13}.  Fluctuations around this ground state,
\be
\label{16}
P_\pm (q,t) - P^0_\pm (q) \equiv \eta_\pm (q,t),
\ee
satisfy the equations of motion
\be
\label{17}
\partial_t \eta_\pm (q,t) = \mp \partial_q \left[P_0 (q) \,\eta_\pm
(q,t) \pm \frac{1}{2} \eta^2_\pm (q,t)\right].
\ee

If the fluctuations are small so that they never cross the asymptotes
$p = \pm q$ of the hyperbola defined by eqn. \eq{3}, then one can
rewrite eqns. \eq{17} in a form that exhibits the presence of a
massless particle.  This is done by introducing the time-of-flight
variable $\tau$, defined by
\be
\label{18}
q \equiv - |2\mu|^{\frac{1}{2}} \cosh \tau, \ \ \ P_0 (q) =
|2\mu|^{\frac{1}{2}} \sinh \tau, \ \ \ 0 \leq \tau < \infty,
\ee
where we have assumed that the fluctuations are confined to the left
half of the hyperbola $(q < 0)$.  We now introduce the new variables
$\bar\eta_\pm (\tau,t)$ defined by
\be
\label{19}
\bar\eta_\pm (\tau,t) \equiv P_0 (q(\tau)) \eta_\pm(q,t).
\ee
They satisfy the equations of motion
\be
\label{20}
(\partial_t \mp \partial_\tau) \bar\eta_\pm (\tau,t) = \partial_\tau
\left[\bar\eta^2_\pm (\tau,t)/2P^2_0(q(\tau))\right].
\ee
Furthermore, one can also deduce the commutation relations
\beq
\label{21}
\left[\bar\eta_\pm (\tau,t),\bar \eta_\pm(\tau',t)\right] &=& \pm 2\pi
i \partial_\tau \delta(\tau-\tau'), \nonumber \\[2mm]
\left[\bar\eta_+(\tau,t),\bar\eta_- (\tau',t)\right] &=& 0,
\eeq
since we know the hamiltonian for the fluctuations
\beq
\label{22}
H_{\rm fluc.} &=& \int \frac{dp \ dq}{2\pi} h(p,q) \delta u(p,q,t)
\nonumber \\[2mm]
&=& \frac{1}{4\pi} \int^\infty_0 d\tau \left[\bar\eta^2_+ (\tau,t) +
\bar\eta^2_-(\tau,t) + \frac{1}{3P^2_0(q(\tau))} \left(\bar\eta^3_+
(\tau,t) - \bar\eta^3_-(\tau,t)\right)\right]
\eeq
Finally, there is the `fixed area' (i.e. fixed fermion number)
constraint eqn. \eq{11}, which reads now
\be
\label{23}
\int^\infty_0 d\tau\left(\bar\eta_+ (\tau,t) -
\bar\eta_-(\tau,t)\right) = 0.
\ee
Equations \eq{19}-\eq{23}\ define the massless scalar field of the $c=1$
matrix model.  One can now obtain the scattering amplitudes and
discuss various other properties of this model.  We refer to a recent
review \cite{polreview}\ for details and original references.

\section{The Leg-Pole Connection with  String Theory}

It has been known for some time now that the tree-level scattering
amplitudes for the matrix model scalar above do not exactly coincide
with the tree-level scattering amplitudes for the tachyon in
2-dimensional string theory \cite{fourteen}.  The difference can be
understood in terms of a wave-function renormalization and is a simple
momentum-dependent phase factor for real momenta.  In coordinate space
this renormalization factor relates the Hilbert space of the matrix
model to that of the string theory by a nonlocal transform of the
states \cite{thirteen}.  Denoting the tachyon field of 2-dimensional
string theory by ${\cal T} (x,t)$ ($x,t$ are space-time labels), this
relationship can be expressed as
\be
\label{24}
{\cal T}_{\rm in} (t+x) = \int^{+\infty}_{-\infty} d\tau \
f\!\left(|\frac{\mu}{2}|^{\frac{1}{2}} e^{\tau-(t+x)}\right)
\bar\eta_{+ \, {\rm in}} (\tau),
\ee
\be
\label{25}
{\cal T}_{\rm out} (t-x) = -\int^{+\infty}_{-\infty} d\tau
\ f\!\left(|\frac{\mu}{2}|^{\frac{1}{2}} e^{-\tau + (t-x)}\right)
\bar\eta_{- \, {\rm out}} (\tau),
\ee
where the `in' and `out' refer, as usual, to the asymptotic fields
obtained in the limits $t \rightarrow -\infty$ and $t \rightarrow
+\infty$ respectively.  In both cases $x$ is taken to be large and
positive, keeping respectively $(t + x)$ and $(t - x)$ fixed.  The
function $f(\alpha)$ is given by
\be
\label{26}
f(\alpha) \equiv \frac{1}{2\sqrt{\pi}} J_0
\left(2\left(\frac{2}{\pi}\right)^{\frac{1}{8}} \sqrt{\alpha}\right),
\ \ \ \alpha \geq 0,
\ee
where $J_0$ is the standard Bessel function of order zero
\cite{twentysix}.

It was recently emphasized in \cite{thirteen} that such a nonlocal
transform is necessary for extracting the space-time physics of
2-dimensional string theory from the matrix model.  In the following
we shall present evidence which not only reinforces this but, among
other things, also seems to suggest that, in fact, it may be possible
to set up a detailed \underbar{operator} correspondence between the
matrix model and string theory by means of a nonlocal and nonlinear
transform.

\section{The Transform -- General Considerations}

Equations \eq{24} and \eq{25} constitute both the starting point and
our motivation for the following enquiry.  We wish to explore
the possibility of a more detailed mapping from the matrix model to
string theory, which would also be valid away from the asymptotic
space-time region of eqs. \eq{24} and \eq{25}.  Moreover, we wish to
develop the general framework \underbar{without} using any specific
parametrization of the fermi fluid boundary fluctuations $\delta
u(p,q,t)$, unlike the collective field $\bar\eta_\pm$ that appear in
eqs. \eq{24} and \eq{25}.  As mentioned earlier, this would enable us
to avoid spurious singularities that might dynamically develop in any
specific (initially nonsingular) parametrization.  We will see an
example of how this works later.  Let us for the moment go on with the
building of the general framework.

Now, the most general form for a matrix model $\leftrightarrow$ string
theory mapping in a perturbative framework must have the expansion
\beq
\label{27}
{\cal T} (x,t) &\equiv& \int dp \ dq \ G_1(x;p,q) \delta u(p,q,t)
\nonumber \\[2mm]
& & + \frac{1}{2} \int dp \ dq \int dp' \ dq' \ G_2(x; p, q; p',q')
\delta u(p,q,t) \delta u(p',q',t) \nonumber \\[2mm]
& & + \cdots
\eeq
The dots represent higher order terms in the fluctuation $\delta u$.
We shall assume that the fluctuations have support only around the
left half of the fermi surface hyperbola, eqn. \eq{3}, i.e. only for
$q \leq - |2\mu|^{\frac{1}{2}}$.  This is consistent with the
perturbative framework within which we are working.  Also, we have
chosen the kernels $G_1 (x;p,q)$, $G_2(x;p,q;p',q')$, etc., to be
time-independent because in the present work we would like to recover
the results of perturbative string theory in flat space and linear
dilaton background.  In this background time evolution in string
theory is controlled by the \underbar{same} hamiltonian as in the
matrix model, namely $H_{\rm fluc.}$ as given in eqn. \eq{22}.

We would now like to ask whether there is a choice of the
kernels $G_1(x;p,q)$, $G_2(x;p,q;p',q')$, etc., which will allow us to
identify the scalar field ${\cal T}(x,t)$, defined by eqn. \eq{27},
with the tachyon of 2-dimensional string theory.  One criterion for
this identification is that ${\cal T}(x,t)$ should satisfy the
$\sigma$-model tachyon $\beta$-function equation in flat space and
linear dilaton background \cite{fifteen,thirteen}:
\be
\label{28}
(\partial^2_t - \partial^2_x) {\cal T}(x,t) = -4g^{-1}_s (2x+c) e^{-2x}
{\cal T}(x,t) + 2\sqrt{2} e^{-2x} {\cal T}^2(x,t) + \cdots
\ee
where the dots represent possible terms of higher order in $e^{-2x}$
as well as higher order in ${\cal T}(x,t)$.  The string coupling $g_s$
and the constant $c$ are defined by
\be
\label{29}
g^{-1}_s \equiv \frac{|\mu|}{\sqrt{2\pi}}, \ \ \ c \equiv 1 + 4\Gamma'
(1) + {\rm ln}g_s.
\ee
The linear term in ${\cal T}(x,t)$ of the specific form appearing on
the r.h.s. of eqn. \eq{28} is known from string scattering amplitudes
and reflects the linear dilaton background \cite{sixteen}.

In addition to eqn. \eq{28}, the operator ${\cal T}(x,t)$ defined by
eqn. \eq{27} must satisfy another crucial criterion for it to be
identified with the tachyon of 2-dimensional string theory.  We must
ensure that the choice of the kernels $G_1,G_2$, etc., that satisfies
eqn. \eq{28} is \underbar{consistent} with the choice that makes
${\cal T}(x,t)$ and its conjugate $\Pi_{\cal T} (x,t)$, defined by
\be
\label{30}
\Pi_{\cal T} (x,t) \equiv -i\left[{\cal T}(x,t),H_{\rm fluc.}\right],
\ee
satisfy the canonical commutation relations:
\beq
\label{31}
\left[{\cal T}(x,t),{\cal T}(y,t)\right] &=& 0, \\[2mm]
\label{32}
\left[{\cal T}(x,t),\Pi_{\cal T}(y,t)\right] &=& i\delta (x-y).
\eeq
Note that the conjugate of ${\cal T}(x,t)$ is defined by eqn. \eq{30}
because, as we mentioned earlier, the generator of time-translations
in the string theory in flat background is identical to that in the
matrix model, namely $H_{\rm fluc.}$ as given in eqn. \eq{22}.

We emphasize that in the present framework eqns. \eq{31} and \eq{32}
are not automatically satisfied even if eqn. \eq{28} is arranged.  The
operator ${\cal T}(x,t)$ inherits a certain quantization from the
matrix model via the r.h.s. of eqn. \eq{27}.  It is not a priori clear
that the \underbar{same} choice of the kernels $G_1,G_2$, etc., that
satisfies the classical equation \eq{28} also necessarily satisfies
the canonical commutation relations in eqns. \eq{31} and \eq{32}.

If there exists a choice of the kernels $G_1,G_2$, etc., order by
order in perturbation theory, such that eqns. \eq{28}, \eq{31} and
\eq{32} are satisfied, then we may identify ${\cal T}(x,t)$ with the
tachyon of 2-dimensional string theory.

Before closing this section, we mention that a choice of the kernels
satisfying the above criteria must necessarily reduce to the
asymptotic form implied in eqns. \eq{24} and \eq{25}, at asymptotic
space-time.  Together with eqn. \eq{28}, this means that corrections
to the kernels, away from the asymptotic region, must be of the
general form
\beq
\label{33}
G_1 (x;p,q) &\buildrel x\to \infty \over \sim&
f(-q e^{-x}) + O(x e^{-2x}) \\[2mm]
\label{34}
G_2, \ G_3, \ {\rm etc.}  &\buildrel x\to \infty \over \sim&
O(e^{-2x})
\eeq
The function $f$ above is the same as that given in eqn. \eq{26}.  The
precise fall-off of the correction terms as we approach the asymptotic
region in eqns. \eq{33} and \eq{34} is dictated by eqn. \eq{28}.

There is an immediate consequence of eqns. \eq{33} and \eq{34}.  Together
with the general form of the matrix model $\leftrightarrow$ string
theory mapping in eqn. \eq{27}, they imply that the tree-level
scattering amplitudes of ${\cal T}(x,t)$ are
\underbar{entirely} determined by the asymptotic kernel function $f$.
To appreciate this point fully we will give below a short derivation
of scattering amplitudes from eqn.
\eq{27}, before we proceed to seek a solution of eqns. \eq{28},
\eq{31} and \eq{32}.  It is useful to do this in any case, since our
derivation of the amplitudes will also illustrate our earlier
assertion that a specific choice of parametrization of the fermi fluid
boundary fluctuations is not required to derive physical properties of
the field ${\cal T}(x,t)$ from the matrix model.

\section{Scattering  Amplitudes}

We will compute the scattering amplitudes at tree-level as usual by
resolving the `out' field ${\cal T}_{\rm out} (t-x)$ in terms of the
`in' field ${\cal T}_{\rm in} (t+x)$.  Since these are defined at
asymptotic times $t \rightarrow \pm \infty$, together with $x
\rightarrow \infty$, where the interactions vanish, it is immediately
clear that the detailed form of the correction terms away from the
asymptotic region in eqns. \eq{33} and \eq{34} will not enter in the
construction of the `out' and `in' fields from eqn. \eq{27}.  It is,
therefore, sufficient for us to write
\be
\label{35}
{\cal T} (x,t) = \int dp \ dq \ f(-q e^{-x}) \delta u(p,q,t) +
O(xe^{-2x})
\ee
for the purposes of computing the tree-level scattering amplitudes.

We will now use a simple trick to shift the time-dependence from the
fluctuation $\delta u(p,q,t)$ to the function $f$.  This is done by
noting that (i) the measure $(dp \, dq)$ for integration over phase
space is invariant under area-preserving diffeomorphisms and (ii)
time-evolution of $\delta u(p,q,t)$ is equivalent to an
area-preserving diffeomorphism on it by the hamiltonian $h(p,q)$ in
eqn. \eq{1}.  In other words, we make the following change of
variables
\be
\label{36}
(p \pm q) e^{\mp t} = (p' \pm q') e^{\mp t'}
\ee
from $(p, q)$ to $(p', q')$ in the integral in eqn. \eq{36}, with $t$
and $t'$ appearing as fixed parameters in the change of variables.
Under this change of variables, the measure $(dp \ dq)$ and the fermi
surface, eqn. \eq{3}, are invariant.  Moreover, using the equation of
motion of $\delta u(p,q,t)$
\be
\label{37}
(\partial_t + p\partial_q + q\partial_p) \delta u(p,q,t) = 0,
\ee
we deduce that
\be
\label{38}
\delta u(p,q,t) = \delta u(p',q',t').
\ee
Therefore, we get from eqn. \eq{35}
\be
\label{39}
{\cal T}(x,t) = \int dp' \ dq' \ f\!\left(-e^{-x} \left[q' \cosh(t - t')
+ p' \sinh(t - t')\right]\right) \delta u (p',q',t') + O(xe^{-2x}).
\ee
The r.h.s. of eqn. \eq{39} is actually independent of $t'$, as can be
easily varified by using eqn. \eq{37}.  Thus, the choice $t' = t$
gives back eqn. \eq{35}.  What we have achieved by rewriting eqn.
\eq{35} in the form of eqn. \eq{39} is to shift the entire
$t$-dependence into the argument of the kernel.  Most importantly,
however, we have in this way introduced a parameter $t'$, which we may
regard as some initial value of time.  The fermi fluid boundary
fluctuation then enters eqn. \eq{39} \underbar{only} as a boundary
condition.  In a more standard notation, writing $t' = t_0$ and
dropping the `primes' from $p$ and $q$, we get
\be
\label{40}
{\cal T} (x,t) = \int dp \ dq \ f\!\left(-e^{-x}\left[q \cosh(t - t_0)
+ p \sinh(t - t_0)\right]\right) \delta u(p,q,t_0) + O(xe^{-2x}).
\ee

Equation \eq{40} proves our assertion that ${\cal T} (x,t)$ is insensitive
to any singularities that specific parametrizations of the fluctuation
$\delta u(p,q,t)$ might dynamically generate even if the parametrization
was perfectly nonsingular to begin with.

To proceed further, we will now use the $t_0$-independence of the
r.h.s. of eqn. \eq{40} to make the choice $t_0 \rightarrow - \infty$.
It is then convenient to use the quadratic profile parametrization for
$\delta u(p,q,t_0)$, which is essentially given by the fields
$\bar\eta_\pm (\tau,t_0)$ discussed in Sec. 1.  We choose the profile
such that in the limit $t_0
\rightarrow - \infty$, $\bar\eta_-$ vanishes. Also, in this limit
$\bar\eta_+ (\tau,t_0) = \bar\eta^0_+ (t_0 + \tau)$ is a function of
$(\tau + t_0)$ only.  Using this, and the formalism developed in eqs.
\eq{12} - \eq{23}, in eqn. \eq{40} in the limit $t_0 \rightarrow -
\infty$, we get
\be
\label{41}
{\cal T} (x,t) = \int^{+\infty}_{-\infty} d\tau \int^{\bar\eta^0_+
(\tau)}_0 d\epsilon \ f\left(\left|\frac{\mu}{2}\right|^{\frac{1}{2}}
\left[e^{\tau -(t + x)} + \left(1 -
\frac{\epsilon}{|\mu|}\right) e^{-\tau + (t-x)}\right]\right) +
O(xe^{-2x}).
\ee

It is now trivial to write down expressions for the `in' and `out'
fields from eqn. \eq{41},
\beq
\label{42}
{\cal T}_{\rm in} (t + x) &=& \int^{+\infty}_{-\infty} d\tau \
f\!\left(\left|\frac{\mu}{2}\right|^{\frac{1}{2}} e^{\tau -
(t+x)}\right) \bar\eta^0_+ (\tau), \\[2mm]
\label{43}
{\cal T}_{\rm out} (t - x) &=& \int^{+\infty}_{-\infty} d\tau
\int^{\bar\eta^0_+ (\tau)}_0 d\epsilon \
f\!\left(\left|\frac{\mu}{2}\right|^{\frac{1}{2}} \left(1 -
\frac{\epsilon}{|\mu|} \right) e^{-\tau + (t-x)}\right),
\eeq
and using these the tree-level amplitudes are easily obtained by
eliminating $\bar\eta^0_+$ and expressing ${\cal T}_{\rm out}$ as a
power series in ${\cal T}_{\rm in}$.  It is a simple exercise to check
that the correct string amplitudes are obtained in this way
\cite{twentyseven}.  We see that the amplitudes are determined
entirely by the asymptotic kernel function $f$.

\section{The Transform -- Specific Form}

We now return to the question of whether there exists an explicit
choice of the kernels $G_1$, $G_2$, etc., that satisfies eqns. \eq{28},
\eq{31} and \eq{32}.

Let us begin with eqn. \eq{28}.  Requiring that ${\cal T}(x,t)$
defined in eqn. \eq{27} satisfy it gives, after a tedious but
straightforward calculation, certain differential equations for the
kernels.  In deriving these differential equations one uses the
equation of motion of the fluctuations, eqn. \eq{37}.  It turns out
that these differential equations can be explicitly solved and
explicit expressions can be obtained for the kernels $G_1$ and $G_2$, these
being the only two kernels relevant to the accuracy of the present
calculations.  Since the calculations are rather straightforward, we
will not give the details here but will merely list the results which
are conveniently summarized in the following parametrization of the
kernels:
\beq
\label{44}
G_1 (x;p,q) &\equiv& f(\alpha) + \left|\frac{\mu}{2}\right| e^{-2x}
\left(2xG(\alpha) + K(\alpha)\right) + O(xe^{-4x}), \\[2mm]
\label{45}
G_2(x;p,q;p',q') &\equiv& e^{-2x} \left[\left(1 +
\frac{pp'}{qq'}\right) F_+(\alpha,\alpha') + \left(1 -
\frac{pp'}{qq'}\right) F_-(\alpha,\alpha')\right] \nonumber \\[2mm]
& & \hspace{2.1in} + \ O(xe^{-4x}),
\eeq
where $\alpha \equiv -  q e^{-x}$ and $\alpha' \equiv - q' e^{-x}$.
The function $f(\alpha)$ is given by eqn. \eq{26} and
\beq
\label{46}
G(\alpha) &\equiv& - \left(\frac{2}{\pi}\right)^{\frac{1}{4}}
f'(\alpha), \\[2mm]
\label{47}
K(\alpha) &\equiv& - \left(\frac{2}{\pi}\right)^{\frac{1}{4}}
\left[(c + 1) f'(\alpha) + \frac{f(\alpha)}{\alpha}\right]
- \frac{f'(\alpha)}{\alpha}, \\[2mm]
\label{48}
F_+ (\alpha,\alpha') &\equiv& \frac{1}{4} f'(\alpha) \delta(\alpha -
\alpha') + \frac{1}{\sqrt{2}}
\left(\frac{2}{\pi}\right)^{-\frac{1}{4}} (\alpha - \alpha')^{-1}
\nonumber \\[2mm]
& & (\alpha \partial_\alpha - \alpha' \partial_{\alpha'})
\left(f(\alpha) f(\alpha')\right), \\[2mm]
\label{49}
F_- (\alpha,\alpha') &\equiv& - \frac{\sqrt{\pi}}{2} f'(\alpha)
f'(\alpha').
\eeq
Here $f'(\alpha) \equiv \frac{d}{d\alpha} f(\alpha)$.  In obtaining
\eq{46}-\eq{49} we have made extensive use of the differential
equation satisfied by $f(\alpha)$,
\be
\label{50}
\left(\alpha f'(\alpha)\right)' = -
\left(\frac{2}{\pi}\right)^{\frac{1}{4}} f(\alpha),
\ee
and the properties of Bessel functions of integer order \cite{twentysix}.

One remarkable thing about the solutions in eqns. \eq{46}-\eq{49} is
that the kernels $G_1$ and $G_2$ are determined entirely in terms of
the asymptotic kernel function $f$.  This is, however, not really
surprising.  The reason for this is that, as we have
seen in the previous section, the information about string scattering
amplitudes is entirely encoded in the function $f$ and
that eqn. \eq{28}, which was used to fix $G_1$ and $G_2$, reproduces
these amplitudes \cite{thirteen}.  The last statement of course
presumes that the canonical commutation relations, eqs. \eq{31} and
\eq{32}, are satisfied.  So let us now turn to these.

We need to check that the operator ${\cal T}(x,t)$ defined by eqns.
\eq{27} and \eq{44}-\eq{49} and its conjugate $\Pi_{\cal T} (x,t)$,
obtained using eqn. \eq{30}, satisfy eqns. \eq{31} and \eq{32}.  In
order to carry out this check it is convenient, though by no means
necessary, to rewrite ${\cal T}(x,t)$ in terms of the collective field
parametrization of the fluctuation $\delta u(p,q,t)$.  In terms of the
collective field variables $\bar\eta_\pm (\tau,t)$ of Sec. 2, ${\cal
T}(x,t)$ is given by


\beq
\label{51}
{\cal T}(x,t) &=& \int^\infty_0 d\tau(\bar\eta_+ - \bar\eta_-) f(v)
\nonumber \\[2mm]
& & + e^{-2x} \Bigg[- \left|\frac{\mu}{2}\right|
\left(\frac{2}{\pi}\right)^{\frac{1}{4}} \int^\infty_0 d\tau
(\bar\eta_+ - \bar\eta_-) \left\{(2x + c + 1) f'(v) +
\frac{f(v)}{v}\right\} \nonumber \\[2mm]
& & \hspace{1.4cm} + \frac{1}{4} \int^\infty_0 d\tau (\bar\eta^2_+ +
\bar\eta^2_-) \frac{f'(v)}{v} - \frac{\sqrt{\pi}}{4} \int^\infty_0 d\tau
\int^\infty_0 d\tau'\nonumber \\[2mm]
& & \hspace{1.4cm} \left\{(\bar\eta_+ - \bar\eta_-) (\bar\eta'_+ -
\bar\eta'_-) - (\bar\eta_+ + \bar\eta_-) (\bar\eta'_+ +
\bar\eta'_-)\right\} f'(v) f'(v') \nonumber \\[2mm]
& & \hspace{1.4cm} + \frac{1}{2\sqrt{2}}
\left(\frac{\pi}{2}\right)^{\frac{1}{4}}
\int^\infty_0 d\tau \int^\infty_0 d\tau' \nonumber \\[2mm]
& & \hspace{1.4cm} \left\{(\bar\eta_+ -
\bar\eta_-) (\bar\eta'_+ - \bar\eta'_-) + (\bar\eta_+ + \bar\eta_-)
(\bar\eta'_+ + \bar\eta'_-)\right\} \times \nonumber \\[2mm]
& & \hspace{2cm} \times (v - v')^{-1} (v\partial_v - v' \partial_{v'})
\left(f(v) f(v')\right)\Bigg] \nonumber \\[2mm]
& & + O(xe^{-4x}, \bar\eta^3_\pm)
\eeq
where $v \equiv \left|\frac{\mu}{2}\right|^{\frac{1}{2}} e^{\tau-x}$,
$v' \equiv \left|\frac{\mu}{2}\right|^{\frac{1}{2}} e^{\tau'-x}$.  We
have also used the short-hand notation, $\bar\eta_\pm \equiv
\bar\eta_\pm (\tau,t)$, $\bar\eta'_\pm \equiv \bar\eta'_\pm
(\tau',t)$, etc.  On the other hand, $f' (v) \equiv \frac{d}{dv}
f(v)$.

Commuting this with the hamiltonian $H_{\rm fluc.}$ given in eqn.
\eq{22} using the commutation relations in eqn. \eq{21}, we get the
following expression for the conjugate of ${\cal T}(x,t)$:

\newpage

\beq
\label{52}
\Pi_{\cal T} (x,t) &\equiv& -i\left[{\cal T}(x,t),H_{\rm fluc.}\right]
\nonumber \\[2mm] &=& -\int^\infty_0 d\tau (\bar\eta_+ + \bar\eta_-)
\partial_\tau f(v) \nonumber \\[2mm]
& & + e^{-2x} \Bigg[\left|\frac{\mu}{2}\right|
\left(\frac{2}{\pi}\right)^{\frac{1}{4}} \int^\infty_0
d\tau(\bar\eta_+ + \bar\eta_-) \nonumber \\[2mm]
& & \hspace{1.3cm} \left\{(2x + c + 1) \partial_\tau
f'(v) + \partial_\tau\left(\frac{f(v)}{v}\right)\right\}
\nonumber \\[2mm]
& & \hspace{1.3cm} + \frac{1}{4} \left(\frac{2}{\pi}\right)^{\frac{1}{4}}
\int^\infty_0 d\tau\left(\bar\eta^2_+ - \bar\eta^2_-\right)
\frac{f(v)}{v} \nonumber \\[2mm]
& & \hspace{1.3cm} + \frac{\sqrt{\pi}}{2} \int^\infty_0 d\tau
\int^\infty_0 d\tau' (\bar\eta_+ + \bar\eta_-) (\bar\eta'_+
-\bar\eta'_-) \times \nonumber \\[2mm]
& & \hspace{1.3cm} \times (v\partial_v - v'\partial_{v'}) (f'(v) f'(v'))
\nonumber \\[2mm]
& & \hspace{1.3cm} - \frac{1}{\sqrt{2}}
\left(\frac{\pi}{2}\right)^{\frac{1}{4}} \int^\infty_0 d\tau
\int^\infty_0 d\tau' (\bar\eta_+ + \bar\eta_-) (\bar\eta'_+ -
\bar\eta'_-)  \times \nonumber \\[2mm]
& & \hspace{1.3cm} \times (v\partial_v + v'\partial_{v'}) (v - v')^{-1}
(v\partial_v - v'\partial_{v'})
(f(v) f(v'))\Bigg] \nonumber \\[2mm]
& & + O(xe^{-4x},\bar\eta^3_\pm).
\eeq

Notice that ${\cal T}(x,t)$ (as well as $\Pi_{\cal T} (x,t)$) contains
\underbar{both} combinations $(\eta_+ - \eta_-)$ and $(\eta_+ +
\eta_-)$.  It is, therefore, not obvious that it would satisfy eqn.
\eq{31}.  A straightforward calculation using the commutation
relations, eqn. \eq{21}, however, shows that eqn. \eq{31} is indeed
satisfied by the combination of terms in eqn. \eq{51}.  This is a
rather nontrivial check on our construction.  We made extensive use of
integrals of two Bessel functions of integer order \cite{twentyeight}
in carrying out this calculation.  Actually, in the quoted
reference the lower limit on the relevant integrals is zero while we
get the lower limit as $e^{-x/2}$ (or $e^{-y/2}$).  The difference,
however, is at least of order $e^{-(x+y)}$, which can safely be
ignored to the accuracy of our present considerations (because there
is already a factor of $e^{-2x}$ (or $e^{-2y}$) present outside the
square brackets in eqn. \eq{51}).

The commutator in eqn. \eq{32} is somewhat more nontrivial to verify.
The subtlety comes from the fact that the first term in eqn. \eq{51}
already gives the required answer for the commutator with the first
term in eqn. \eq{52}.  So the contribution of the rest of the terms to
the commutator is required to vanish, but it is not clear how the
contribution of terms linear in $x$ (or $y$) would cancel since, on
the face of it, there is only one such contribution.  Explicit
calculation, however, shows that the last term in the curly brackets
in eqn. \eq{51} (and eqn. \eq{52}) which is linear in $\bar\eta_\pm$
gives a contribution that is proportional to the integral
\[
\int^\infty_{e^{-x/2}} \frac{d\nu}{\nu} J_0 (\nu) J_0\left(\nu
e^{\frac{x-y}{2}}\right).
\]
Because of a logarithmic divergence we can now not naively set the
lower limit of integration to zero, as was done previously.  The
divergence can be extracted by rewriting this integral as
\[
\int^\infty_{e^{-x/2}} \frac{d\nu}{\nu} J_0 (\nu) \left[J_0 \left(\nu
e^{\frac{x-y}{2}}\right) - 1\right] + \int^\infty_{e^{-x/2}}
\frac{d\nu}{\nu} J_0 (\nu).
\]
We may now safely set the lower limit in the first term to zero.  The
second term can be evaluated explicitly in the limit $x \rightarrow
\infty$.  The logarithmic divergence appears as a \underbar{linear}
term in $x$.  This precisely cancels the contribution to the
commutator coming from terms in eqns. \eq{51} and \eq{52} which are
manifestly linear in $x$.  It is remarkable that eqn. \eq{32} is
satisfied in this rather nontrivial fashion.

Now that our construction of ${\cal T} (x,t)$ has passed all the
consistency checks, we may identify it with the tachyon of
2-dimensional string theory.

\section{Discrete States and Other Backgrounds}

One of the aspects of 2-dimensional string theory most difficult to
understand in the matrix model has been the apparent absence of any
trace of the so-called $W^-$ discrete states \cite{six}.  This issue
is rather crucial since, among other things, these states contain the
dilaton black hole.  The first hint that a nonlocal transform is
necessary to extract the physics of discrete states from the matrix
model was seen in the gravitational scattering calculation of
\cite{thirteen}.  Now that we have seen strong evidence that it might,
in fact, even be possible to set up a detailed operator correspondence
between string theory and matrix model by a nonlocal and nonlinear
transform, it has become easy to at least begin to imagine how the
$W^-$ discrete states might make their appearance in this setting.  It
seems natural to think that these states correspond to
\underbar{other} nonlocal and nonlinear combinations of the fluid
boundary fluctuation $\delta u(p,q,t)$, which satisfy properties
expected of the $W^-$ discrete operators.  Included among these
properties must be the following \cite{six}.  These operators must
commute with each other and with the tachyon ${\cal T}(x,t)$ and they
must transform as a representation of the $W$-infinity algebra
generated by the operators given in eqn. \eq{8}.  (We will have more
to say about this last point in the next section.)

A motivation for believing that discrete states, like the tachyon,
correspond to nonlocal and nonlinear combinations of the density
fluctuations comes from the full set of string $\sigma$-model
$\beta$-function equations:
\beq
\label{53}
R_{\mu\nu} + 2\nabla_\mu\nabla_\nu \Phi - 2\partial_\mu T \partial_\nu
T &=& 0, \nonumber \\[2mm]
R + 4\nabla^2\Phi - 4(\nabla \Phi)^2 - 2(\nabla T)^2 - 8T^2 &=& 16,
\nonumber \\[2mm]
\nabla^2 T - 2\nabla \Phi \nabla T - 4T - 2\sqrt{2} T^2 &=& 0.
\eeq
Let us fix the gauge in which the dilaton is always given (at least in
a local neighbourhood) by its linear form, namely,
\[
\Phi = - 2x,
\]
and the metric has the diagonal form
\[
g_{\mu\nu} = {\rm diag} \, (g_{00},g_{11}).
\]
Perturbing around the flat background
\[
g_{00} = 1 + g_0, \ \ \ g_{11} = -1 + g_1,
\]
gives the following equations for the metric fluctuations:
\beq
\label{54}
\partial_x g_0 &=&  {1 \over 2}
\left[ (\partial_t T)^2 + (\partial_x T)^2 + 4 T^2 \right] -4 g_1,
\nonumber \\[2mm]
\left( \partial_x +4 \right) g_1 &=&
- {1\over 2} \left[(\partial_t T)^2 + (\partial_x T)^2 - 4 T^2 \right]
\nonumber \\[2mm]
\partial_t g_1 &=& - (\partial_t T) (\partial_x T).
\eeq

In the absence of the tachyon field, eqns. \eq{54} lead to the famous
black-hole solution \cite{nine} for the metric perturbations
\[
g_0 = g_1 \propto e^{-4x}.
\]
With the tachyon field present,
eqns. \eq{54} are the
analogues of eqn. \eq{28} for the present case of the metric
perturbations.  Knowing $T(x,t)$ $\left(= e^{-2x} \{ {\cal T}(x,t) -
\frac{1}{\sqrt{2} g_s} (x + c)\}\right)$ in terms of the matrix
model variables, for example as given in \eq{51}, allows
us to rewrite these equations in terms of the density fluctuations of
the matrix model.  In this way, one obtains \underbar{new} nonlocal
and nonlinear combinations of the matrix model variables which
represent the metric fluctuations.

It seems reasonable to assume that this holds true for the higher
tensor modes of the string as well.  If that is so, then we may be
able to construct all the discrete state operator combinations, even
without knowing their equations.  This is because all the $W^-$
discrete state operators must together form a representation of the
$W$-infinity algebra \cite{six}.  It seems clear to us that it is the
techniques of $W$-infinity representation theory that will then prove
to be more powerful and more useful (than any higher order
$\sigma$-model calculations to incorporate the higher tensor modes of
the string theory) for discovering these operators in the matrix
model.

Connected with the question of discrete states is the issue of
backgrounds other than the flat space and linear dilaton background we
have considered here.  These backgrounds can be created by using
coherent state techniques once we know the discrete state operators
constructed along the above lines.  The tachyon operator in eqn.
\eq{51} would, in general, have a nonzero value in such coherent
states.  One effect of this would be to change the asymptotic kernel
in the transform to string theory.  It would, therefore, seem that
different backgrounds should correspond to different choices of the
asymptotic kernel.  This line of enquiry, as well as the construction
of discrete state operators outlined above, is under active
investigation.

\section{Realization of $W_\infty$-Symmetry}

Given that it is the nonlocal operator ${\cal T}(x,t)$ in eqn.
\eq{51} (and not $\bar\eta_\pm$ themselves) that has physical
(space-time) significance, it is of interest to ask how the generators
of $W$-infinity symmetry act on it.  This question is actually also of
interest in view of the discussion in the previous section.  Moreover,
the answer to this question should, in particular, also give us an
infinite number of symmetries of the equation of motion \eq{28}.

It turns out that the action of $W$-infinity symmetry on ${\cal T}
(x,t)$ is \underbar{nonlocal}, as one might have expected.  Since this
is easiest to see in the collective field parametrization, let us
consider an example in this parametrization to illustrate the above
statement.  But first note that the $W$-infinity generators are
represented on the fluctuations by
\be
\label{56}
\omega_{mn} = e^{-(m-n)t} \int \frac{dp \ dq}{2\pi} (-p-q)^m (p-q)^n
\delta u(p,q,t).
\ee

The example we will consider is that of the `half' of Virasoro, which
is generated by
\be
\label{57}
V_n = \frac{i}{2^{n+1}} \left(\frac{\pi}{2}\right)^{-\frac{n}{4}}
\left[\omega_{n+1,1} + 2\mu \, \omega_{n,0}\right], \ \ \ n \geq 0.
\ee
The $V_n$'s have the following expressions in terms of the collective
variables $\bar\eta_\pm$:
\be
\label{58}
V_n = \frac{g^{-n/2}_s}{4\pi i} \int^\infty_0 d\tau
\left[e^{-n(t+\tau)} \bar\eta^2_+ + e^{-n(t-\tau)}\bar\eta^2_-\right]
+ O(\bar\eta^3_\pm)
\ee
where $g_s$ is the string coupling given by eqn. \eq{29}.  Using the
commutation relations in eqn. \eq{21} one can easily check that
\be
\label{59}
\left[V_n,V_m\right] = (n-m) V_{n+m}.
\ee
In writing eqn. \eq{59} we have ignored the $O(\bar\eta^3_\pm)$ terms
in eqn. \eq{58}.  These will not affect the linearized variations of
the tachyon considered below.  These latter can be obtained from the
linearized variations of the collective variables:
\be
\label{60}
\left[V_n,\bar\eta_\pm (\tau,t)\right] = \mp g_s^{-n/2} \partial_\tau
\left(e^{-n(t\pm \tau)} \bar\eta_\pm (\tau,t)\right) +
O(\bar\eta^2_\pm).
\ee
Using this in eqn. \eq{51}, we get
\be
\label{61}
\left[V_n,(\partial_t - \partial_x) {\cal T}(x,t)\right] =
\int^{+\infty}_{-\infty} dy \ \Delta_n^-(t - x; t - y) (\partial_t -
\partial_y) {\cal T}(y,t) + O({\cal T}^2, {\rm higher\,order\,in\,e^{-x}})
\ee
where
\be
\label{62}
\Delta_n^- (x;y) = -(-)^n \int \frac{dk}{2\pi} e^{ikx} e^{-(ik+n)y}
\frac{\Gamma(1+ik+n) \Gamma(ik+n)}{\Gamma(ik) \Gamma(ik)}.
\ee
For the other branch, we get
\be
\label{621}
\left[V_n,(\partial_t + \partial_x) {\cal T}(x,t)\right] =
\int^{+\infty}_{-\infty} dy \ \Delta_n^+(t + x; t + y) (\partial_t +
\partial_y) {\cal T}(y,t) + O({\cal T}^2, {\rm higher\,order\,in\,e^{-x}})
\ee
where
\be
\label{622}
\Delta_n^+ (x;y) = (-)^n g_s^{-n}
\int \frac{dk}{2\pi} e^{-ikx} e^{(ik - n)y}
\frac{\Gamma(1+ik-n) \Gamma(ik-n)}{\Gamma(ik) \Gamma(ik)}.
\ee

The tachyon transformation law derived in eqns. \eq{61} and \eq{62}
above is precisely the one obtained earlier in \cite{seven} using the
techniques of BRST cohomolgy in liouville string theory (see eqn.
(5.21) of this reference\footnote{Note that to get their result from \eq{61}
and \eq{62} we need to use euclidean momenta.}).  The more standard form
for the Virasoro transformation obtained in this reference (in eqn.
(5.23)) for a redefined `tachyon' is nothing but the transformation in
eqn. \eq{60} for the collective field, which we now know is
\underbar{not} the tachyon of 2-dimensional string theory.  The string
theory tachyon, as we have seen, has a \underbar{nonlocal}
transformation under the Virasoro, and indeed under the full set of
$W$-infinity transformations in eqn. \eq{56}.

There are perhaps at least two reasons for this nonlocal realization
of the $W$-infinity symmetry of the matrix model in string theory.
One is the fact that in the present framework there are no fields
explicitly present which represent the higher tensor modes of the
string (these have implicitly been integrated out).  The other reason
probably is that the $W$-infinity symmetry is the `global' remnant of
a large local symmetry after gauge-fixing.  It is clearly very
important to have a deeper understanding of both these aspects in the
present framework in order to get a better handle on the underlying
structure of the string theory.

\section{Concluding Remarks}

To sum up, we have derived the tachyon $\sigma$-model $\beta$-function
equation of 2-dimensional string theory in flat space and linear
dilaton background, working entirely within the $c=1$ matrix model.
This equation is derived for a nonlocal and nonlinear combination of
the matrix model variables.  We have also seen that the $W$-infinity
symmetry of the matrix model has a nonlocal action on the tachyon
field defined in this way, a result which was known earlier in
liouville string theory.  These results, among other things, present
strong evidence for the viewpoint that the space-time properties of
2-dimensional string theory can only be extracted from the $c=1$
matrix model by means of a nonlocal and nonlinear mapping.  We have
argued that this viewpoint also has the potential to accommodate
discrete states corresponding to the higher tensor modes of the
string.  Some evidence for this already exists \cite{thirteen}, but
one can now hope to even explicitly construct the corresponding
operators.

Throughout this work our considerations have been perturbative.  It is
clear that we need a nonperturbative understanding of the issues
discussed here, if we are to eventually use the full nonperturbative
power of the matrix model to understand some of the stringy issues,
such as the nature of quantum gravity in the strong coupling regime.
Progress on the nonperturbative aspects of the present work is,
therefore, of urgent interest.  Even at the perturbative level,
however, the picture is not completely clear yet.  It is likely that a
better understanding of the emergence of discrete states and the role
of $W$-infinity symmetry at the perturbative level will give us a
better handle on the underlying structure of the space-time theory
and, therefore, probably also on its nonperturbative aspects.  These
issues are under active investigation.

\newpage

\def\und{\underbar}
\def\bib{\bibitem}

\end{document}